\documentclass[review]{elsarticle}

\usepackage{hyperref}
\usepackage{amsmath,graphicx}
\usepackage{textcomp}
\usepackage{psfrag}
\usepackage[printonlyused]{acronym}
\usepackage{lipsum}
\usepackage{booktabs} 
\usepackage{multirow}
\usepackage[tight]{subfigure}
\usepackage{pgfplots}
\usetikzlibrary{patterns}
\usepgfplotslibrary{units}

\pgfplotsset{compat=1.3}
\pgfplotsset{
	/pgfplots/xbar legend/.style={
		/pgfplots/legend image code/.code={%
			\draw[##1,/tikz/.cd, bar width=3pt,yshift=-0.1em,bar shift=0pt]
			plot coordinates {(0.8em, 0cm) (0.6em, 1.7*\pgfplotbarwidth)};},
	}
}

\journal{Elsevier SI: Multi-Microphone ASR}

\begin{document}
	
	
	\newcounter{MYtempeqncnt}
	
	\acrodef{ASR}{Automatic Speech Recognition}
	\acrodef{CDR}{Coherent-to-Diffuse Power Ratio}
	\acrodef{CHiME-3}{3rd CHiME Speech Separation and Recognition Challenge}
	\acrodef{DFT}{Discrete Fourier Transform}
	\acrodef{DNN}{Deep Neural Network}
	\acrodef{DOA}{Direction-of-Arrival}
	\acrodef{FFT}{Fast Fourier Transform}
	\acrodef{fMLLR}{feature-space Maximum Likelihood Linear Regression}
	\acrodef{FIR}{Finite Impulse Response}
	\acrodef{GMM}{Gaussian Mixture Model}
	\acrodef{GCC-PHAT}{Generalized Cross-Correlation with Phase Transform}
	\acrodef{GPU}{Graphics Processing Unit}
	\acrodef{HMM}{Hidden Markov Model}
	\acrodef{LS}{Least-Squares}
	\acrodef{MAPSSWE}{Matched-Pair Sentence-Segment Word Error}	
	\acrodef{MFCC}{Mel-Frequency Cepstral Coefficient}
	\acrodef{MMSE}{Minimum Mean Square Error}
	\acrodef{MVDR}{Minimum Variance Distortionless Response}
	\acrodef{NIST}{National Institute of Standards}
	\acrodef{PDF}{Probability Density Function}
	\acrodef{PSD}{Power Spectral Density}
	\acrodef{PSDs}{Power Spectral Densities}
	\acrodef{SNR}{Signal-to-Noise Ratio}
	\acrodef{RNN}{Recurrent Neural Network}
	\acrodef{STFT}{short-time Fourier transform}
	\acrodef{STFT-UP}{STFT uncertainty propagation}
	\acrodef{TDOA}{Time Difference of Arrival}
	\acrodef{WDS}{Weighted Delay-and-Sum}
	\acrodef{WER}{Word Error Rate}
	\acrodef{WNG}{White Noise Gain}
	\newcommand{\bb}[1]{\mathbf{#1}}
	\newcommand{\mc}[1]{\mathcal{#1}}
	\newcommand{\sine}[1]{\ensuremath{\mathrm{sin}(#1)}}
	\newcommand{\cosine}[1]{\ensuremath{\mathrm{cos}(#1)}}
	\newcommand{\sinc}[1]{\ensuremath{\mathrm{sinc}(#1)}}
	\renewcommand\Re{\operatorname{Re}}
	
	\begin{frontmatter}
		
		\title{Robust coherence-based spectral enhancement for speech recognition in adverse real-world environments}
		
		\author{Hendrik Barfuss, Christian Huemmer, Andreas Schwarz\footnote{A. Schwarz was with the Friedrich-Alexander University of Erlangen-N\"urnberg while the work has been conducted. He is now with Amazon Development Center, Aachen, Germany.}, \\and Walter Kellermann}
		\address{Multimedia Communications and Signal Processing,\\Friedrich-Alexander University Erlangen-N\"urnberg, \\ Cauerstr. 7, 91058 Erlangen, Germany \\ \{barfuss,huemmer,schwarz,wk\}@lnt.de\\[-1mm]}
		
		%
		%
		
		\begin{abstract}
			Speech recognition in adverse real-world environments is highly affected by reverberation and non-stationary background noise. A well-known strategy to reduce such undesired signal components in multi-microphone scenarios is spatial filtering of the microphone signals. In this article, we demonstrate that an additional coherence-based postfilter, which is applied to the beamformer output signal to remove diffuse interference components from the latter, is an effective means to further improve the recognition accuracy of modern deep learning speech recognition systems. To this end, the \ac{CHiME-3} baseline speech enhancement system is extended by a coherence-based postfilter and the postfilter's impact on the \acp{WER} of a state-of-the-art automatic speech recognition system is investigated for the realistic noisy environments provided by \ac{CHiME-3}. To determine the time- and frequency-dependent postfilter gains, we use \ac{DOA}-dependent and \ac{DOA}-independent estimators of the coherent-to-diffuse power ratio as an approximation of the short-time signal-to-noise ratio. Our experiments show that incorporating coherence-based postfiltering into the \ac{CHiME-3} baseline speech enhancement system leads to a significant reduction of the \acp{WER}, with relative improvements of up to $11.31\%$.
		\end{abstract}
		
		\begin{keyword}
			\texttt{Robust speech recognition, Postfiltering, Spectral enhancement, Coherence-to-diffuse power ratio, Wiener filter}
		\end{keyword}
		
	\end{frontmatter}
	
	\acresetall
	
	\section{Introduction}
	\label{sec:intro}
	
	For a satisfying user experience of human-machine interfaces it is crucial to ensure a high accuracy in automatically recognizing the user's speech. However, as soon as no close-talking microphone is used for capturing the desired speech signal, the recognition accuracy suffers from additional reverberation, background noise and active interfering speakers which are picked up by the microphones~\cite{delcroix_2013,Yoshioka_2015}. Techniques for robust speech recognition in such reverberant and noisy environments can be categorized into either front-end (e.g., speech enhancement \cite{cohen_2003,krueger_2010,gales_2011}) or back-end (e.g., acoustic-model adaptation \cite{li_2006,liao_2013,yu_2013}) processing techniques.
	
	The \ac{CHiME-3} \cite{Barker_Chime3} targets the performance of state-of-the-art \ac{ASR} systems in real-world scenarios. The primary goal is to improve the \ac{ASR} performance of real recorded speech of a person talking to a tablet device in realistic everyday noisy environments by employing front-end and/or back-end signal processing techniques.
	To this end, a baseline \ac{ASR} system has been initially provided and updated as follow-up of \ac{CHiME-3} to achieve a high recognition accuracy in everyday real-world scenarios. Front-end processing of the updated baseline now employs the BeamformIt toolkit \cite{Anguera_TASLP_2007} for processing the recorded microphone signals by a \ac{WDS} beamforming technique. The beamformer output is used as input to the \ac{ASR} back-end system which contains a \ac{DNN}-based acoustic model and a \ac{RNN}-based language model.
	
	In this article, we extend the updated \ac{CHiME-3} baseline system by a low-complexity coherence-based postfilter which is applied to the beamformer output signal to further remove reverberation and non-stationary background noise from the latter. The postfilter is realized as a Wiener filter, where, in contrast to the classical Wiener filter, see, e.g., \cite{Diethorn_SNR_2000}, we use an estimate of the \ac{CDR} as an approximation of the short-time \ac{SNR} to compute the time- and frequency-dependent Wiener filter gains. The \ac{CDR}, which is the ratio of the power of direct and diffuse signal components, needs to be estimated from the microphone signals. 
	We compare and evaluate two \ac{DOA}-independent and two \ac{DOA}-dependent \ac{CDR} estimators. Two of the evaluated \ac{CDR} estimators have been proposed and shown to be very effective for dereverberation by Schwarz and Kellermann in \cite{lnt2014-28, lnt2015-17}. The remaining two \ac{CDR} estimators have been proposed by Jeub et al.~\cite{jeub_blind_2011} and Thiergart et al.~\cite{thiergart_signal--reverberant_2012,thiergart_spatial_2012} earlier than \cite{lnt2014-28,lnt2015-17}, and are evaluated as reference methods.
	In contrast to the previous work in \cite{lnt2014-28, lnt2015-17}, where the dereverberation performance was evaluated using \acp{WER} of a \ac{HMM}-\ac{GMM}-based \ac{ASR} system trained on clean speech, we now evaluate the efficacy of the \ac{CDR}-based Wiener filter realizations with a state-of-the-art \ac{HMM}-\ac{DNN}-based \ac{ASR} system trained on noisy training data from different acoustic environments (provided by \ac{CHiME-3} \cite{Barker_Chime3}).
	Moreover, the new \ac{CDR} estimators in \cite{lnt2014-28, lnt2015-17} were proposed and evaluated for a dual-channel microphone array, whereas the recognition task of \ac{CHiME-3} involves signal enhancement using a six-channel microphone array. We therefore extended the \ac{CDR} estimation procedure to a multi-channel (here: six-channel) scenario.
	To summarize, the contributions of this article are as follows:
	\begin{enumerate}
		\item First-time evaluation of new \cite{lnt2014-28, lnt2015-17} and previously known \cite{jeub_blind_2011, thiergart_signal--reverberant_2012, thiergart_spatial_2012} \ac{CDR} estimators with a state-of-the-art \ac{HMM}-\ac{DNN}-based \ac{ASR} system in challenging acoustic scenarios.
		\item First-time application of coherence-based dereverberation using the new \ac{CDR} estimators \cite{lnt2014-28, lnt2015-17} to a multi-microphone scenario with more than two microphones.
	\end{enumerate}
	
\begin{figure*}[t]
		\scriptsize
		\centering
		\psfrag{A}[c][c]{STFT}
		\psfrag{B1}[c][c]{Beamformer}
		\psfrag{B2}[c][c]{Subsection~\ref{subsec:MVDR}}
		\psfrag{C1}[c][c]{Postfilter}
		\psfrag{C2}[c][c]{Subsection~\ref{subsec:CDRpostfilter}}
		\psfrag{D1}[c][c]{ASR back-end system}
		\psfrag{D2}[c][c]{Section~\ref{sec:BackEnd}}
		\psfrag{E}[c][c]{Recogn.}
		\includegraphics[scale=.7]{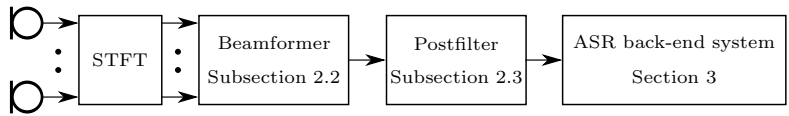}
		\caption{Overview of the overall signal processing pipeline system with beamformer and postfilter as acoustic front-end signal processing. The acoustic back-end system, including feature extraction/transformation, is equal to the updated baseline acoustic back-end system of \ac{CHiME-3} (see Section~\ref{sec:BackEnd} for more details on the employed \ac{ASR} system).}
		\label{fig:ASR_pipe}
	\end{figure*}

	An overview of the signal processing pipeline employed in this work is given in Figure~\ref{fig:ASR_pipe}. While the purpose of the beamformer is to spatially focus on the target source, i.e., to reduce the signal components from interfering point sources, the postfilter shall remove diffuse interference components, e.g., reverberation, from the beamformer output signal. The output of the front-end signal enhancement (consisting of beamformer and postfilter) is further processed by the \ac{ASR} back-end system, which provides an \ac{HMM}-\ac{DNN}-based speech recognizer (see Section~\ref{sec:BackEnd}). 
	
	The remainder of this article is structured as follows: In Section~\ref{sec:FrontEnd}, the employed front-end signal enhancement is described in detail, followed by a presentation of the employed \ac{ASR} system in Section~\ref{sec:BackEnd}. Results of our evaluation are presented in Section~\ref{sec:Exper}. There, the performance of the front-end speech enhancement using four different \ac{CDR} estimators is evaluated with respect to \acp{WER} of the \ac{ASR} back-end system. Furthermore, we analyze the statistical significance of the obtained \ac{WER} improvements. Section~\ref{sec:summary_conclusion} provides concluding remarks.

	\section{Front-end enhancement techniques}
	\label{sec:FrontEnd}
	The front-end speech enhancement considered in this article consists of a \ac{WDS} beamformer (based on the BeamformIt toolkit \cite{Anguera_TASLP_2007}) and a single-channel coherence-based Wiener filter. In the following, we introduce the signal model which is used throughout this paper. Then, the baseline \ac{WDS} beamformer is briefly reviewed, followed by a detailed presentation of the coherence-based Wiener filter based on \ac{DOA}-independent and \ac{DOA}-dependent \ac{CDR} estimation.
	
	\subsection{Signal model}
	For a consistent presentation of the front-end speech enhancement, we first introduce a signal model which will be used throughout this article.
	
	The $N$ microphone signals of the microphone array in the \ac{STFT} domain at frame $l$ and frequency $f$ are given as:
	\begin{equation}
	\bb{x}(l,f) = \bb{h}(l,f) S(l,f) + \bb{n}(l,f),
	\label{eq:signalModel}
	\end{equation}
	where vector
	\begin{equation}
	\bb{x}(l,f) = [X_0(l,f),\, X_1(l,f),\, \ldots,\, X_{N-1}(l,f)]^{T}
	\label{eq:signalModel_X}
	\end{equation}
	contains the microphone signals, $S(l,f)$ denotes the clean source signal, and 
	$\bb{n}(l,f)$ includes sensor noise as well as interference and diffuse background noise components and is defined analogously to $\bb{x}(l,f)$ in (\ref{eq:signalModel_X}). Assuming free-field propagation of sound waves, $\bb{h}(l,f)$ represents the steering vector modeling the sound propagation of a plane wave arriving from the target source \ac{DOA} $(\phi_\mathrm{d},\theta_\mathrm{d})$ (azimuth and elevation angle $\phi$ and $\theta$ are measured with respect to the positive $x$- and $z$-axis, respectively, as in \cite{VanTrees:2004}) to all $N$ microphones:
	\begin{equation}
	\bb{h}(l,f) = [e^{j 2 \pi f \tau_{0}(l)},\, e^{j 2 \pi f \tau_{1}(l)}, \, \ldots,\, e^{j 2 \pi f \tau_{N-1}(l)}]^{T},
	\label{eq:signalModel_H}
	\end{equation}
	where $\tau_{n}(l)$ is the (possibly time-varying) \ac{TDOA} for the target source direction of the $n$-th channel with respect to the origin of the coordinate system \cite{VanTrees:2004}:
	\begin{equation}
	\tau_{n}(l) = \frac{\bb{a}^{T}(l) \bb{p}_{n}}{c}.
	\label{eq:signalModel_TDOA}
	\end{equation}
	In (\ref{eq:signalModel_TDOA}), vector $\mathbf{p}_{n}$ contains the position of the $n$-th microphone in Cartesian coordinates and the unit vector $\bb{a}(l)$ is defined as \cite{VanTrees:2004}:
	\begin{equation}
	\bb{a}(l) = - [ \sine{\theta_\mathrm{d}}\cosine{\phi_\mathrm{d}},\, \sine{\theta_\mathrm{d}}\sine{\phi_\mathrm{d}},\, \cosine{\theta_\mathrm{d}} ]^{T}.
	\label{eq:signalModel_directionVector}
	\end{equation}
	Note that $\bb{a}(l)$ points from the target source \ac{DOA} towards the origin of the coordinate system (hence the minus sign) and that the time-dependency of $\phi_\mathrm{d}$ and $\theta_\mathrm{d}$ in (\ref{eq:signalModel_directionVector}) has been omitted for brevity. Moreover, $c$ represents the speed of sound and operator $(\cdot)^T$ denotes the transpose of a vector or matrix.
	%
	The beamformer output $Y_\mathrm{BF}(l,f)$ is obtained by multiplying each microphone signal with a complex-valued beamformer coefficient $W_{n}(l,f)$, followed by a summation over all microphone channels:
	\begin{equation}
	Y_\mathrm{BF}(l,f) = \bb{w}^{T}(l,f)\bb{x}(l,f),
	\end{equation}
	where 
	\begin{equation}
	\bb{w}(l,f) = [W_{0}(l,f), \ldots, W_{N-1}(l,f)]^{T}
	\label{eq:BFweights_vector}
	\end{equation}
	contains the beamformer coefficients $W_{n}(l,f)$.
	
	Subsequently, the postfilter is applied to the beamformer output signal, yielding the overall output signal
	\begin{equation}
	Y(l,f) = G(l,f)Y_\mathrm{BF}(l,f),
	\end{equation}
	where $G(l,f)$ describes the postfilter gain at frame $l$ and frequency $f$. The enhanced signal $Y(l,f)$ is used as input of the acoustic back-end system (described in Section~\ref{sec:BackEnd}).
	
	\subsection{Weighted delay-and-sum beamformer}
	\label{subsec:MVDR}
	
	The employed beamformer is provided by the BeamformIt toolkit \cite{Anguera_TASLP_2007} and based on the \ac{WDS} beamforming technique \cite{VanVeen_IEEE_ASSP_1998}. The $n$-th beamformer filter coefficient $W_{n}(l,f)$ at frame $l$ and frequency $f$ is given as
	\begin{equation}
	W_{n}(l,f) =  w_{n}(l) e^{- j 2 \pi f \tau_{n}(l)},
	\label{eq:W_BF_main}
	\end{equation}
	where $w_{n}(l)$ is the frequency-independent weight for the $n$-th channel. By modifying $W_{n}(l,f)$, one can control the shape and angular direction of the beamformer's main beam.
	
	The \acp{TDOA} are estimated using the \ac{GCC-PHAT} localization technique, see, e.g., \cite{Brandstein_ICASSP_1997}. Before the signals are time-aligned a two-step post processing is applied to the estimated \acp{TDOA}: First, a noise threshold is estimated and employed to remove non-reliable \ac{TDOA} estimates which may have been obtained from non-speech or noisy segments. Second, Viterbi decoding of the remaining \ac{TDOA} values is performed to maximize the speaker continuity, i.e., to avoid steering the beam to noise sources which are only present at a very short time span.
	
	The channel weights are chosen adaptively over time, starting with the classical delay-and-sum as initial value: $W_{n}(0,f) = 1/N$. Furthermore, automatic channel selection and elimination is performed to avoid using microphone signals of poor quality. Both, channel weight adaptation and channel selection/elimination are based on the cross-correlations between the microphone channels.
	
	For a more detailed explanation of the baseline beamformer (including localization), as provided by the BeamformIt toolkit, we refer the reader to \cite{Anguera_TASLP_2007, Anguera_PHDThesis_2006}.
	
	\subsection{Coherence-based postfilter}
	\label{subsec:CDRpostfilter}
	As illustrated in Figure~\ref{fig:ASR_pipe}, we apply a postfilter to remove diffuse noise components from the beamformer output signal. The postfilter gain $G(l,f)$ at frame $l$ and frequency $f$ is given as \cite{Diethorn_SNR_2000, haensler_acoustic_2004}:
	\begin{equation}
	G(l,f) = \mathrm{max} \left\{ 1 - \mu \frac{1}{1 + \mathrm{SNR}(l,f)} ,\, G_\mathrm{min} \right\},
	\label{eq:WienerFilter}
	\end{equation}
	with overestimation factor $\mu$, and gain floor $G_\mathrm{min}$. The postfilter in (\ref{eq:WienerFilter}) is a Wiener filter which uses the short-time \ac{SNR} to compute the filter gain $G(l,f)$. In this work, we approximate the short-time \ac{SNR} in (\ref{eq:WienerFilter}) by an estimate of the so-called \ac{CDR}, which is the power ratio between direct and diffuse signal components. From (\ref{eq:WienerFilter}), it can be seen that a low \ac{CDR} value, which corresponds to strong diffuse signal components being present at the input of the system, leads to low filter gains and vice versa.
	
	The \ac{CDR} between two omnidirectional microphones is defined as \cite{lnt2014-28}:
	\begin{equation}
	\mathrm{CDR}(l,f) = \frac{\Gamma_\mathrm{n}(l,f) - \Gamma_\mathrm{x}(l,f)}{\Gamma_\mathrm{x}(l,f)-\Gamma_\mathrm{s}(l,f)},
	\label{eq:CDRgeneral}
	\end{equation}
	where $\Gamma_\mathrm{x}(l,f)$, $\Gamma_\mathrm{s}(l,f)$, $\Gamma_\mathrm{n}(l,f)$ denote the coherence functions of the observations, of the direct-path signal, and of the noise between two observation points (microphones), respectively.
	In the following, the two microphones are indexed by the variables $p=1,...,N$ and $q=1,...,N$, respectively.
	To this end, the spatial coherence functions for the direct-path signals and diffuse noise components are given as
	\begin{align}
	\Gamma_\mathrm{s}(l,f) &= e^{j 2 \pi f (\tau_{p}(l) -\tau_{q}(l)) },\\
	\Gamma_\mathrm{n}(f) &= \frac{\sin(2 \pi f \frac{d_{pq}}{c})}{2 \pi f \frac{d_{pq}}{c}},
	\label{eq:GammaDiff}
	\end{align}
	respectively, with \ac{TDOA}s $\tau_{p}(l)$, $\tau_{q}(l)$ calculated in (\ref{eq:signalModel_TDOA}) and microphone spacing $d_{pq}$. Moreover, a short-time estimate $\hat{\Gamma}_\mathrm{x}(l,f)$ of the coherence function of both microphone signals $\Gamma_\mathrm{x}(l,f)$ in (\ref{eq:CDRgeneral}) can be obtained using
	\begin{equation}
	\hat{\Gamma}_\mathrm{x}(l,f) = \frac{\hat{\Phi}_{x_px_q}(l,f) }{\sqrt{\hat{\Phi}_{x_px_p}(l,f)\hat{\Phi}_{x_qx_q}(l,f)}}
	\label{eq:CoherenceEst}
	\end{equation}
	by estimating the auto- and cross-\ac{PSDs} $\hat{\Phi}_{x_px_q}(l,f)$ from the microphone signals $X_p(l,f)$ and  $X_q(l,f)$ based on recursive averaging
	\begin{equation}
	\hat{\Phi}_{x_px_q}(l,f) = \lambda \hat{\Phi}_{x_px_q}(l-1,f) + (1-\lambda) X_p(l,f) X^*_q(l,f)
	\label{eq:RecAv}
	\end{equation}
	with forgetting factor $\lambda$. Operator $(\cdot)^*$ creates the conjugate complex of $(\cdot)$.
	However, inserting the coherence estimate $\hat{\Gamma}_\mathrm{x}(l,f)$ into~(\ref{eq:CDRgeneral}) is not feasible due to the mismatch between coherence models and actual acoustic conditions as $\mathrm{CDR}(l,f)$ might become a complex-valued quantity \cite{lnt2015-17}.
	Thus, the \ac{CDR} needs to be estimated. In this work, we use two \ac{CDR} estimators, which have been proposed and shown to be especially effective in \cite{lnt2014-28,lnt2015-17}, and which are given by 
	{\small
		\begin{align}
		& \widehat{\mathrm{CDR}}_\text{DOAindep} = \notag \\ &  \max \left(0, \frac{\Gamma_\mathrm{n}\, \Re\{\hat\Gamma_\mathrm{x}\} -{|\hat\Gamma_\mathrm{x}|}^2 - \sqrt{\Gamma_\mathrm{n}^2\, {\Re\{\hat\Gamma_\mathrm{x}\}}^2 - \Gamma_\mathrm{n}^2\, {|\hat\Gamma_\mathrm{x}|}^2 + \Gamma_\mathrm{n}^2 - 2\, \Gamma_\mathrm{n}\, \Re\{\hat\Gamma_\mathrm{x}\} + {|\hat\Gamma_\mathrm{x}|}^2}}{{|\hat\Gamma_\mathrm{x}|}^2 - 1} \right)
		\label{eq:CDR_estimator_prop2}
		\end{align}
	}
	and
	\begin{equation}
	\widehat{\mathrm{CDR}}_\text{DOAdep} =  \max \left(0, \frac{1-\Gamma_\mathrm{n} \cos(\arg(\Gamma_\mathrm{s}))}{|\Gamma_\mathrm{n} - \Gamma_\mathrm{s}|} \left| \frac{\Gamma_\mathrm{s}^* (\Gamma_\mathrm{n}-\hat\Gamma_\mathrm{x}) }{\Re \{ \Gamma_\mathrm{s}^* \hat\Gamma_\mathrm{x}\}-1} \right| \right),
	\label{eq:CDR_estimator_prop3}
	\end{equation}
	respectively, where $\Re\{\cdot\}$, $|\cdot|$, and $\arg\{\cdot\}$ represent the real part, magnitude, and phase of $(\cdot)$, respectively. The maximum operation is required to prevent negative results for the \ac{CDR} estimate. Note that frame- and frequency-index have been omitted in (\ref{eq:CDR_estimator_prop2}) and (\ref{eq:CDR_estimator_prop3}) for brevity.
	As can be seen from (\ref{eq:CDR_estimator_prop2}), this estimator does not require the \ac{TDOA} of the target source, since $\Gamma_{s}(l,f)$ is not required for calculating $\widehat{\mathrm{CDR}}_\text{DOAindep}$. This is not the case for the \ac{DOA}-dependent estimator $\widehat{\mathrm{CDR}}_\text{DOAdep}$ in (\ref{eq:CDR_estimator_prop3}).
	On the one hand, the advantage of \ac{DOA}-independent estimators is that these are easier to realize, since no additional information, i.e., the \ac{DOA} (or \ac{TDOA}) of the target source, is required. On the other hand, \ac{DOA}-dependent estimators have a directional filtering effect due to the incorporation of the \ac{DOA} into the \ac{CDR} estimate, which is expected to provide a more accurate \ac{CDR} estimate and, therefore, lead to a better suppression of diffuse noise and reverberation.
	
	In addition to (\ref{eq:CDR_estimator_prop2}) and (\ref{eq:CDR_estimator_prop3}), we evaluate two \ac{CDR} estimators as reference methods, which have been chosen due to their good speech dereverberation performance in previous experiments \cite{lnt2014-28,lnt2015-17}.			
	
	The first reference \ac{CDR} estimator is a \ac{DOA}-independent estimator and was proposed by Thiergart et al.~in \cite{thiergart_signal--reverberant_2012,thiergart_spatial_2012}:
	\begin{equation}
	\widehat{\mathrm{CDR}}_\text{Thiergart} =  \max \left(0, \Re \left\{ \frac{\Gamma_\mathrm{n}-\hat\Gamma_\mathrm{x}}{\hat\Gamma_\mathrm{x}-e^{j \arg \hat\Gamma_\mathrm{x}}} \right\} \right).
	\label{eq:CDR_estimator_thiergart2}
	\end{equation}
	This estimator is \ac{DOA}-independent, because it uses the instantaneous phase of the coherence estimate $\hat\Gamma_\mathrm{x}(l,f)$ as phase estimate for the direct signal model, i.e., $\hat\Gamma_\mathrm{s}(l,f) = e^{j \arg \hat\Gamma_\mathrm{x}(l,f)}$. 
	
	The second reference \ac{CDR} estimator is based on Jeub's \ac{CDR} estimator \cite{jeub_blind_2011}. The estimator is given as:
	\begin{equation}
	\widehat{\mathrm{CDR}}_\text{Jeub} =  \max \left(0, \frac{\Gamma_\mathrm{n}-\Re \{\Gamma_\mathrm{s}^* \hat\Gamma_\mathrm{x}\}}{\Re \{\Gamma_\mathrm{s}^* \hat\Gamma_\mathrm{x}\}-1} \right)
	\label{eq:CDR_estimator_jeub}
	\end{equation}
	and is \ac{DOA}-dependent. Jeub et al.~relied on the same assumption as McCowan and Bourlard, who derived a Wiener postfilter for a coherent signal in diffuse noise \cite{mccowan_microphone_2003}, to explicitly formulate their \ac{CDR} estimate (\ref{eq:CDR_estimator_jeub}). Hence, (\ref{eq:CDR_estimator_jeub}) can also be derived using McCowan's and Bourlard's signal and noise \acs{PSD} estimates \cite{schwarz:daga2014}.
	
	In \cite{lnt2014-28,lnt2015-17} it was shown that for non-zero \acp{TDOA} both $\widehat{\mathrm{CDR}}_\text{Thiergart}$ and $\widehat{\mathrm{CDR}}_\text{Jeub}$ provide biased \ac{CDR} estimates, whereas for \acp{TDOA} equal to zero, their \ac{CDR} estimate is unbiased. A more detailed investigation of the employed \ac{CDR} estimators (\ref{eq:CDR_estimator_prop2})--(\ref{eq:CDR_estimator_jeub}) with respect to their properties and dereverberation performance for a two-channel microphone array using an \ac{HMM}-\ac{GMM}-based \ac{ASR} system (trained on clean speech) and signal-dependent measures, can be found in \cite{lnt2014-28, lnt2015-17}.

	When applying the coherence-based postfilter to the output of a beamformer, two aspects need to be considered: First, since the microphone array of the \ac{CHiME-3} challenge consists of five forward-facing microphones (in this work, the additional backward-facing microphone was excluded from the experiments), the \ac{CDR} estimator (initially designed for a pair of microphones) has to be adapted to exploit all available microphone signals. To do so, we apply the \ac{CDR} estimators in (\ref{eq:CDR_estimator_prop2})--(\ref{eq:CDR_estimator_jeub}) to every pair of non-failing microphones (detection and elimination of failing channels is carried out by the BeamformIt toolkit, see Subsection \ref{subsec:MVDR}) to obtain the \ac{CDR} estimate for each microphone pair.
	
	From each of the \ac{CDR} estimates, we calculate the respective diffuseness values as \cite{lnt2015-17, galdo_diffuse_2012}
	\begin{equation}
	\mathrm{D}(l,f) = (1 + \widehat{\mathrm{CDR}}(l,f))^{-1}.
	\label{eq:diffuseness}
	\end{equation}
	Subsequently, we take the arithmetic mean of all microphone pair-specific diffuseness values to obtain an average diffuseness estimate $\overline{\mathrm{D}}(l,f)$. We then
	calculate the final \ac{CDR} estimate as:
	\begin{equation}
	\widehat{\mathrm{CDR}}_\mathrm{In}(l,f) = \frac{1 - \overline{\mathrm{D}}(l,f)}{\overline{\mathrm{D}}(l,f)}.
	\label{eq:CDRestimate_input}
	\end{equation}
	We take the average of the diffuseness estimates instead of averaging the microphone pair-specific \ac{CDR} values directly, since the latter can take values between zero and infinity, whereas the diffuseness lies in the interval $0 \leq \mathrm{D}(l,f) \leq 1$. Note that McCowan and Bourlard also applied their Wiener postfilter to a multi-microphone setup \cite{mccowan_microphone_2003}. 
	%
	The second aspect is that $\widehat{\mathrm{CDR}}_\mathrm{In}(l,f)$ is a \ac{CDR} estimate at the input of the signal enhancement system, i.e., at the microphones. However, what we need is the \ac{CDR} at the output of the beamformer. This can be obtained by applying a correction factor $A_\mathrm{\Gamma}(l,f)$ to $\widehat{\mathrm{CDR}}_\mathrm{In}(l,f)$. Thus, the \ac{CDR} estimate at the output of the beamformer $\widehat{\mathrm{CDR}}_\mathrm{BF}(l,f)$ is defined as
	\begin{equation}
	\widehat{\mathrm{CDR}}_\mathrm{BF}(l,f) = \frac{\widehat{\mathrm{CDR}}_\mathrm{In}(l,f)}{ A_\mathrm{\Gamma}(l,f) },
	\label{eq:CDR_bf}
	\end{equation}
	where $A_\mathrm{\Gamma}(l,f)$ is the inverse of the array gain for diffuse noise, given by \cite{simmer_post-filtering_2001}
	\begin{equation}
	A_\mathrm{\Gamma}(l,f) = \bb{w}^{H}(l,f) \mathbf{J}_\mathrm{diff}(f) \bb{w}(l,f),
	\label{eq:A}
	\end{equation}
	where $(\cdot)^H$ denotes the Hermitian of a vector or matrix and $\mathbf{J}_\mathrm{diff}(f)$ is the $N \times N$ spatial coherence matrix of a diffuse noise field with the ($p,q$)-th element given by (\ref{eq:GammaDiff}). Note that we assume a distortionless beamformer response for the target source \ac{DOA}. Hence, the denominator of $A_\mathrm{\Gamma}(l,f)$ in (\ref{eq:A}) is equal to one.
	
\begin{figure}[t]
		\centering
		\scriptsize
		\psfrag{A}[c][c]{STFT}
		\psfrag{B1}[c][c]{Beamformer}
		\psfrag{C1}[c][c]{Coherence}
		\psfrag{C2}[c][c]{estimation}
		\psfrag{D1}[c][c]{CDR}
		\psfrag{D2}[c][c]{estimation}
		\psfrag{X1}[cl][cl]{$X_{0}(l,f)$}
		\psfrag{XN}[cl][cl]{$X_{N-1}(l,f)$}
		\psfrag{Y_bf}[c][c]{$Y_\mathrm{BF}(l,f)$}
		\psfrag{Y}[cl][cl]{$Y(l,f)$}
		\psfrag{G}[c][c]{\parbox{3cm}{\centering Spectral enhancement\\[-1mm]$G(l,f)$}}
		\psfrag{CDR_in}[cl][cl]{$\widehat{\mathrm{CDR}}_\mathrm{In}(l,f)$}
		\psfrag{CDR_bfout}[cl][cl]{$\widehat{\mathrm{CDR}}_\mathrm{BF}(l,f)$}
		\psfrag{Ainv}[c][c]{$1/A_\mathrm{\Gamma}(l,f)$}
		\psfrag{Coh estimation}[c][c]{\parbox{1.5cm}{\centering Coherence estimation}}
		\psfrag{Beamforming}[c][c]{\parbox{2cm}{\centering Beamforming}}
		\psfrag{CDR estimation}[c][c]{\parbox{2cm}{\centering CDR\\ estimation}}
		\psfrag{Gamma}[c][c]{$\hat{\Gamma}_\mathrm{x}(l,f)$}
		\includegraphics[width=0.9\columnwidth]{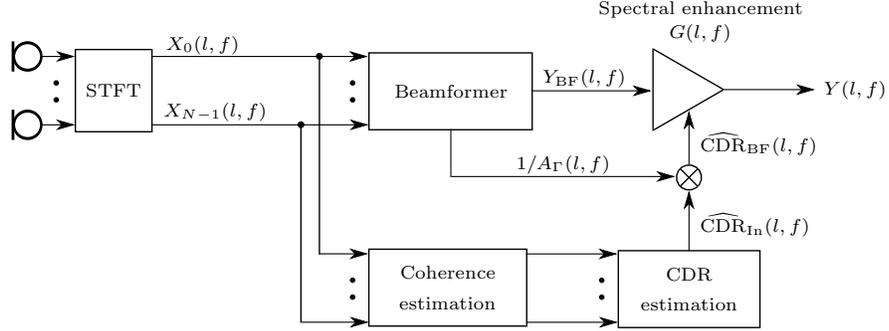}
		\caption{Illustration of the front-end signal processing consisting of beamformer and coherence-based postfilter, where the latter is applied to the beamformer output signal.}
		\label{fig:CDRbased_postfilter}
	\end{figure}

	Figure~\ref{fig:CDRbased_postfilter} shows the block diagram of the employed front-end enhancement system, consisting of beamformer and coherence-based postfilter. We would like to point out that knowledge of the array geometry is required to estimate the target source \acp{TDOA} which are needed to realize the beamformer (\ref{eq:W_BF_main}) and the Wiener filter (if one of the \ac{DOA}-dependent \ac{CDR} estimators in (\ref{eq:CDR_estimator_prop3}) or (\ref{eq:CDR_estimator_jeub}) is to be used for estimating the \ac{CDR}).

	\section{ASR back-end system}
	\label{sec:BackEnd}
	In this section, we describe the employed \ac{ASR} back-end system, which was updated by the organizers of \ac{CHiME-3} based on the findings of the challenge results.
	The employed \ac{ASR} system includes an \ac{HMM}-\ac{GMM} system, consisting of $2500$ tied triphone \ac{HMM} states which are modeled by $15000$ Gaussians, as well as an \ac{HMM}-\ac{DNN} system providing state-of-the-art \ac{ASR} performance for real-world scenarios. The \ac{HMM}-\ac{DNN} system employs a seven-layer feed-forward \ac{DNN} with $2048$ neurons per hidden layer and is based on ``Karel's implementation`` of the Kaldi toolkit~\cite{povey_2011}.
	The \ac{DNN} training process is using noisy data from different acoustic environments and includes pre-training using restricted Boltzmann machines, cross entropy training, and sequence discriminative training using the state-level minimum Bayes risk (sMBR) criterion.
	We extract $40$ \acp{MFCC} from the single-channel output signal of the acoustic front-end signal enhancement scheme. The resulting length-$40$ feature vector is passed through per-speaker mean and variance normalization, a \ac{fMLLR} transformation and context extension, where the latter appends the feature vectors of $3$ (HMM-GMM system) or $5$ (HMM-DNN system) previous and successive frames.
	The CHiME-3 baseline system performs acoustic front-end signal enhancement during training and decoding (i.e., we train on enhanced features obtained from the output signals of the acoustic front-end including beamforming and postfiltering).
	In the decoding phase, we produce word lattices ($3$-gram language model), generate a $100$-best list ($5$-gram language model) and perform lattice rescoring using an \ac{RNN}-based language model (one hidden layer with $300$ nodes) trained on the WSJ0 text corpus. More details about the \ac{RNN}-based language modeling can be found in~\cite{hori_asru2015}.

	\section{Experiments}
	\label{sec:Exper}
	In the following, we evaluate the signal enhancement performance of our proposed front-end (see Figure~\ref{fig:CDRbased_postfilter}) consisting of \ac{WDS} beamforming and coherence-based postfiltering. First, we give an overview over the evaluation setup and the choice of front-end parameters. After this, we illustrate the impact of the proposed front-end enhancement on the \ac{STFT} spectra of a noisy speech utterance. 
	Finally, we evaluate the speech recognition accuracy achieved by the \ac{HMM}-\ac{DNN}-based \ac{ASR} system presented in Section~\ref{sec:BackEnd} using our proposed signal enhancement described in Section~\ref{sec:FrontEnd}, and verify the statistical significance of the obtained \ac{WER} improvements.
	
	\subsection{Setup and parameters}
	To obtain the \ac{STFT}-representation, we use a \ac{DFT}-based uniform filterbank with window length $1024$, \ac{FFT} size $512$, and downsampling factor $128$ \cite{Harteneck_TACSII_1999}. The signals were processed at a sampling rate of $16$ kHz.
	\ac{DOA} estimation and \ac{WDS} beamforming were provided by the BeamformIt toolkit. The window and scroll sizes used for BeamformIt were the default baseline parameters, i.e., $500$ms and $250$ms, respectively. The array geometry was known a priori. 
	
	For realizing the coherence-based postfilter, we chose a gain floor $G_\mathrm{min}= 0.1$ and optimized the overestimation factor $\mu$ on the development data set (see Figure~\ref{fig:ResL}). The short-time coherence estimates $\hat{\Gamma}_\mathrm{x}(l,f)$ were obtained by recursive averaging of the auto- and cross-\ac{PSDs} with forgetting factor $\lambda = 0.68$ in (\ref{eq:RecAv}).
	
	The \ac{ASR} task included sets of real and simulated noisy utterances in four different environments: caf\'e (CAF), street junction (STR), public transport (BUS), and pedestrian area (PED). For each environment, a training set, a development set, and an evaluation set consisting of real and simulated data were provided by \ac{CHiME-3} \cite{Barker_Chime3}. The training data set consists of $1600$ real and $7138$ simulated utterances from a total of four (real), and 83 (simulated) speakers. The development data set contains $1640$ real and simulated utterances, whereas the evaluation data set consists of $1320$ real and simulated utterances. The utterances within each data set were obtained from different speakers. In this work, we focused on evaluating the recognition accuracy of the \ac{DNN}-based \ac{ASR} system for the practically relevant case of real-world recordings.

	\subsection{Illustration of front-end impact in the STFT domain}
\begin{figure}
		\centering
		\subfigure{
			\begin{tikzpicture}[scale=1,trim axis left]
			\node at (10.1,2.525) {\scriptsize $[\text{dB}]$};
			\node at (-0.95,2.25) {\small (a)};
			\begin{axis}[
			label style = {font=\scriptsize},
			tick label style = {font=\scriptsize},   
			ylabel style={yshift=-1mm},
			xlabel style={yshift=1mm},
			width=\textwidth,height=3.85cm,grid=major,grid style = {dotted,black},  		
			axis on top,
			enlargelimits=false,
			xmin=1, xmax=245, ymin=0, ymax=8000,	
			xtick={1,50,100,150,200},
			xticklabels={\empty},
			change y base=true,  y SI prefix=kilo,
			ytick={0, 2000, 4000, 6000, 8000},
			ylabel={$f / \text{kHz}\rightarrow$},
			colorbar horizontal, colormap/jet, 
			colorbar style={
				at={(0,1.175)}, anchor=north west, font=\scriptsize, width=0.8\textwidth, height=0.25cm, xticklabel pos=upper
			},
			point meta min=-120, point meta max=0]
			\addplot graphics [xmin=0, xmax=245, ymin=0, ymax=8025] {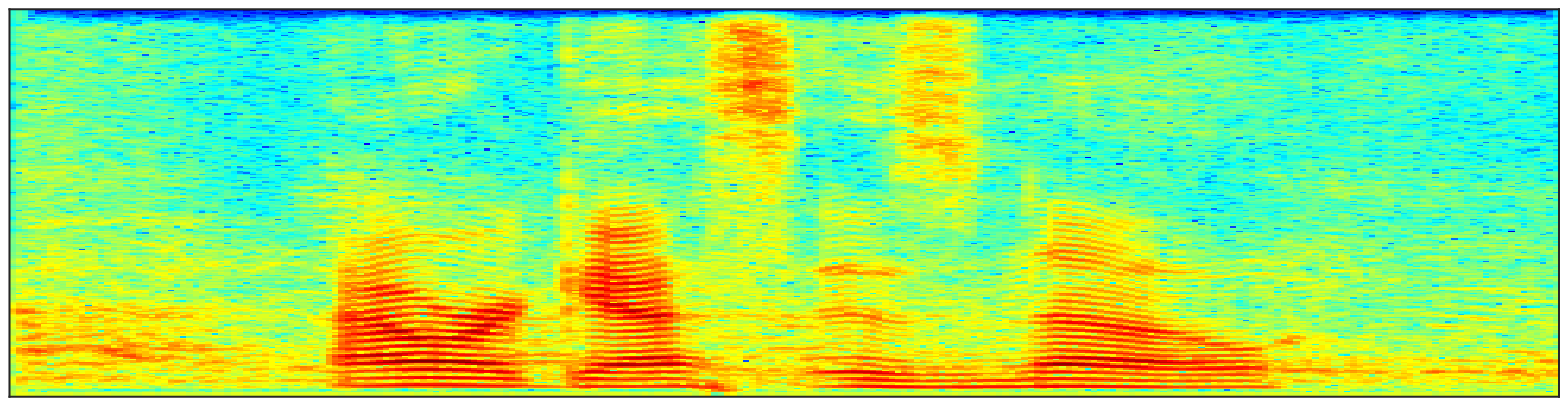};	
			\end{axis}
			\label{fig:illustration_spectrograms_S}
			\end{tikzpicture}
		}\\[-5mm] 
		\subfigure{
			\begin{tikzpicture}[scale=1,trim axis left]
			\node at (-0.95,2.25) {\small (b)};
			\begin{axis}[
			label style = {font=\scriptsize},
			tick label style = {font=\scriptsize}, 
			xlabel style={yshift=1mm},
			ylabel style={yshift=-1mm},  	 
			width=\textwidth,height=3.85cm,grid=major,grid style = {dotted,black},
			axis on top, 	
			enlargelimits=false,
			xmin=1, xmax=245, ymin=0, ymax=8000,	
			xtick={1,50,100,150,200},
			xticklabels={\empty},
			change y base=true,  y SI prefix=kilo,
			ytick={0, 2000, 4000, 6000, 8000},	
			ylabel={$f / \text{kHz}\rightarrow$}]
			\addplot graphics [xmin=0, xmax=245, ymin=0, ymax=8025] {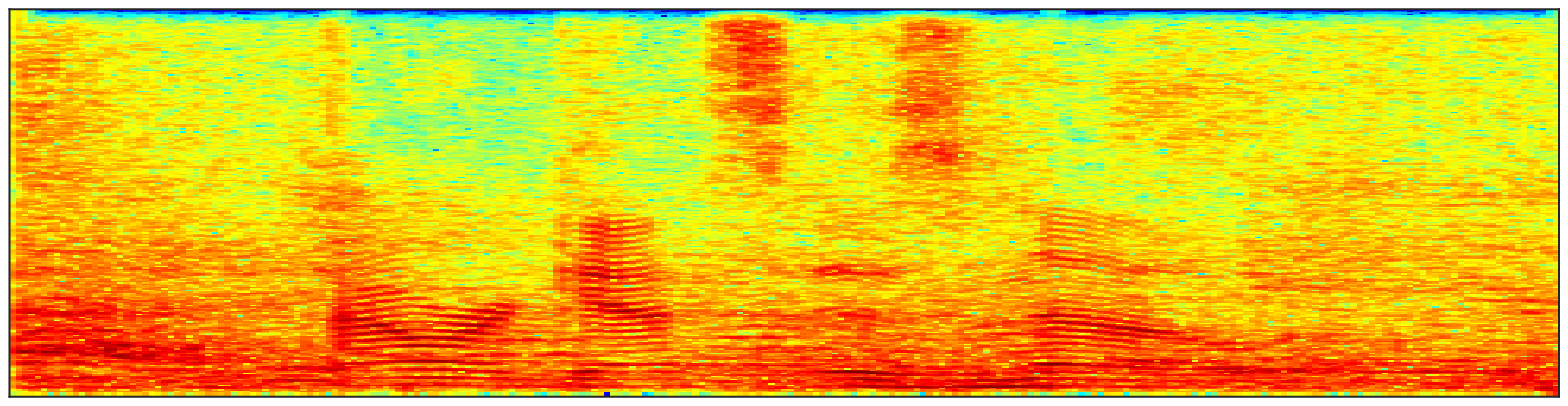};
			\end{axis}
			\label{fig:illustration_spectrograms_X}
			\end{tikzpicture} 
		}\\[-5mm]
		\subfigure{
			\begin{tikzpicture}[scale=1,trim axis left]
			\node at (-0.95,2.25) {\small (c)};
			\begin{axis}[
			label style = {font=\scriptsize},
			tick label style = {font=\scriptsize}, 
			xlabel style={yshift=1mm},
			ylabel style={yshift=-1mm},  	 
			width=\textwidth,height=3.85cm,grid=major,grid style = {dotted,black},
			axis on top, 	
			enlargelimits=false,
			xmin=1, xmax=245, ymin=0, ymax=8000,	
			xtick={1,50,100,150,200},
			xticklabels={\empty},
			change y base=true,  y SI prefix=kilo,
			ytick={0, 2000, 4000, 6000, 8000},	
			ylabel={$f / \text{kHz}\rightarrow$}]
			\addplot graphics [xmin=0, xmax=245, ymin=0, ymax=8025] {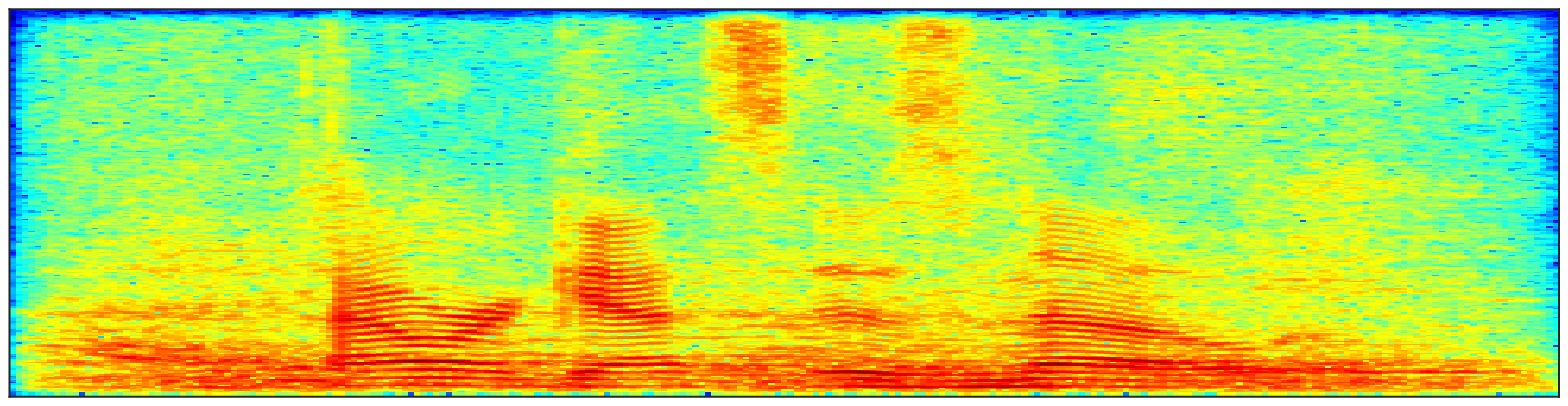};
			\end{axis}
			\label{fig:illustration_spectrograms_YBF}
			\end{tikzpicture} 
		}\\[-5mm]
		\subfigure{
			\begin{tikzpicture}[scale=1,trim axis left]
			\node at (-0.95,2.25) {\small (d)};
			\begin{axis}[
			label style = {font=\scriptsize},
			tick label style = {font=\scriptsize},
			xlabel style={yshift=1mm},
			ylabel style={yshift=-1mm},  	 
			width=\textwidth,height=3.85cm,grid=major,grid style = {dotted,black},
			axis on top, 	
			enlargelimits=false,
			xmin=1, xmax=245, ymin=0, ymax=8000,	
			xtick={1,50,100,150,200},
			xlabel={$l\rightarrow$},
			change y base=true,  y SI prefix=kilo,
			ytick={0, 2000, 4000, 6000, 8000},	
			ylabel={$f / \text{kHz}\rightarrow$}]
			\addplot graphics [xmin=0, xmax=245, ymin=0, ymax=8025] {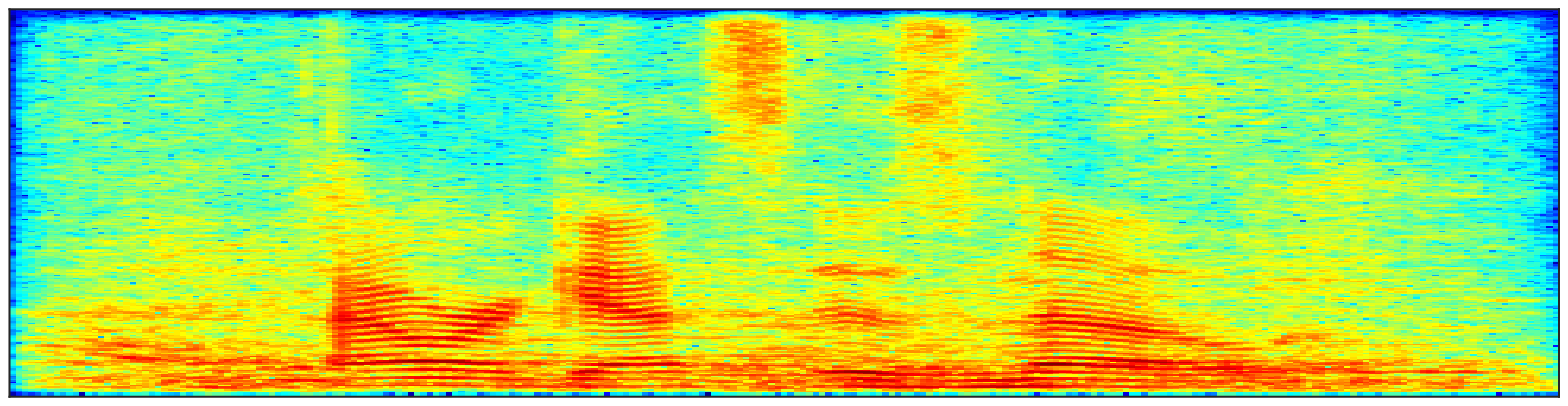};
			\end{axis}
			\label{fig:illustration_spectrograms_Y}
			\end{tikzpicture} 
		}\\[-3mm]
		\subfigure{
			\begin{tikzpicture}[scale=1,trim axis left]  
			\node at (-0.95,2.25) {\small (e)};
			\begin{axis}[
			label style = {font=\scriptsize},
			tick label style = {font=\scriptsize},   
			xlabel style={yshift=1mm},
			ylabel style={yshift=-1mm},
			width=\textwidth,height=3.85cm,grid=major,grid style = {dotted,black},
			axis on top, 	
			enlargelimits=false,
			xmin=1, xmax=245, ymin=0, ymax=8000,	
			xtick={1,50,100,150,200},
			xlabel={$l\rightarrow$},
			change y base=true,  y SI prefix=kilo,
			ytick={0, 2000, 4000, 6000, 8000},
			ylabel={$f/\text{kHz} \rightarrow$},
			colorbar horizontal, colormap/jet, 
			colorbar style={
				at={(0,1.175)}, anchor=north west, font=\scriptsize, width=0.87\textwidth, height=0.25cm, xticklabel pos=upper
			},
			point meta min=0, point meta max=1]
			\addplot graphics [xmin=0, xmax=246, ymin=0, ymax=8025] {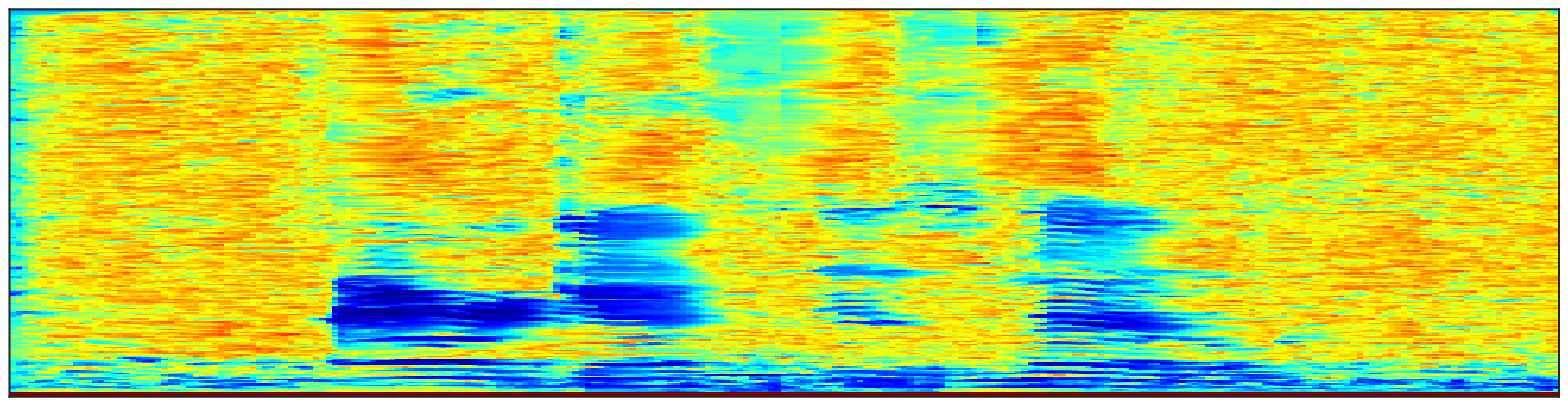};
			\end{axis}
			\label{fig:illustration_spectrograms_D}
			\end{tikzpicture}
		}
		\caption{Illustration of the impact of front-end signal processing on the recorded noisy microphone signal. Spectrograms of (a) desired signal $20\log_{10}(|S(l,f)|)$, recorded by a close-talking microphone, of (b) microphone signal $20\log_{10}(|X_{1}(l,f)|)$, of (c) baseline beamformer output signal $20\log_{10}(|Y_\mathrm{BF}(l,f)|)$, and of (d) postfilter output signal $20\log_{10}(|Y(l,f)|)$ based on the \ac{DOA}-dependent \ac{CDR} estimator in (\ref{eq:CDR_estimator_prop3}). Figure (e) shows the average diffuseness $\overline{\mathrm{D}}(l,f)$, estimated from the microphone signals.}
		\label{fig:illustration_spectrograms}
	\end{figure}

	In Figure~\ref{fig:illustration_spectrograms}, an exemplary illustration of the impact of our proposed front-end, including \ac{WDS} beamformer and coherence-based postfilter with overestimation factor $\mu=1.3$ (at this point, $\mu$ was chosen according to \cite{lnt2015-17}), on the \ac{STFT} spectra of a noisy utterance is shown, with frame $l$ and frequency $f$ on the horizontal and vertical axis, respectively. The coherence-based postfilter was realized using the \ac{DOA}-dependent \ac{CDR} estimator in (\ref{eq:CDR_estimator_prop3}). As a reference signal, the spectrum of the close-talking microphone (channel $0$) is shown in Figure~\ref{fig:illustration_spectrograms_S}. It contains the desired speech signal plus little background noise. The desired signal is a male speaker saying ``\textit{Our guess is no}'' in a caf\'e environment. The spectrum of microphone channel $1$ is illustrated in Figure~\ref{fig:illustration_spectrograms_X}. As can be seen, low- as well as high-frequency noise is acquired by the microphone, whereas most of the noise is present in the frequency range of speech.
	Applying the baseline beamformer already leads to a reduction of the interfering components, as illustrated in Figure~\ref{fig:illustration_spectrograms_YBF}.
	A thorough comparison of Figure~\ref{fig:illustration_spectrograms_YBF} with Figure~\ref{fig:illustration_spectrograms_Y} shows that applying the coherence-based postfilter to the beamformer output further reduces the interference across the entire frequency range.
	The estimated diffuseness $\overline{\mathrm{D}}(l,f)$ at the microphones is illustrated in Figure~\ref{fig:illustration_spectrograms_D}. Comparing Figure~\ref{fig:illustration_spectrograms_D} with the spectrogram of the reference signal in \ref{fig:illustration_spectrograms_S} shows that $\overline{\mathrm{D}}(l,f)$ exhibits low values whenever the target source is active and a comparison to \ref{fig:illustration_spectrograms_YBF} and \ref{fig:illustration_spectrograms_Y} shows that the dereverberation effect is most prominent where diffuseness is high. Furthermore, it can be seen that $\overline{\mathrm{D}}(l,f)$ is large whenever the target source is not active, indicating a high level of diffuse noise components.
	A final comparison of Figures~\ref{fig:illustration_spectrograms_S} and \ref{fig:illustration_spectrograms_Y} reveals the similarity between the front-end output signal $Y(l,f)$ and the close-talking microphone signal $S(l,f)$, which indicates the effectiveness of the proposed front-end signal enhancement technique.
	
\begin{figure}
		\centering
		\subfigure{
			\begin{tikzpicture}[scale=1,trim axis left]
			\node at (10.1,2.525) {\scriptsize $[\text{dB}]$};
			\node at (-0.95,2.25) {\small (a)};
			\begin{axis}[
			label style = {font=\scriptsize},
			tick label style = {font=\scriptsize},   
			ylabel style={yshift=-1mm},
			xlabel style={yshift=1mm},
			width=\textwidth,height=3.85cm,grid=major,grid style = {dotted,black},  		
			axis on top,
			enlargelimits=false,
			xmin=1, xmax=245, ymin=0, ymax=8000,	
			xtick={1,50,100,150,200},
			xlabel={$l\rightarrow$},
			change y base=true,  y SI prefix=kilo,
			ytick={0, 2000, 4000, 6000, 8000},
			ylabel={$f / \text{kHz}\rightarrow$},
			colorbar horizontal, colormap/jet, 
			colorbar style={
				at={(0,1.175)}, anchor=north west, font=\scriptsize, width=0.8\textwidth, height=0.25cm, xticklabel pos=upper
			},
			point meta min=-120, point meta max=0]
			\addplot graphics [xmin=0, xmax=245, ymin=0, ymax=8025] {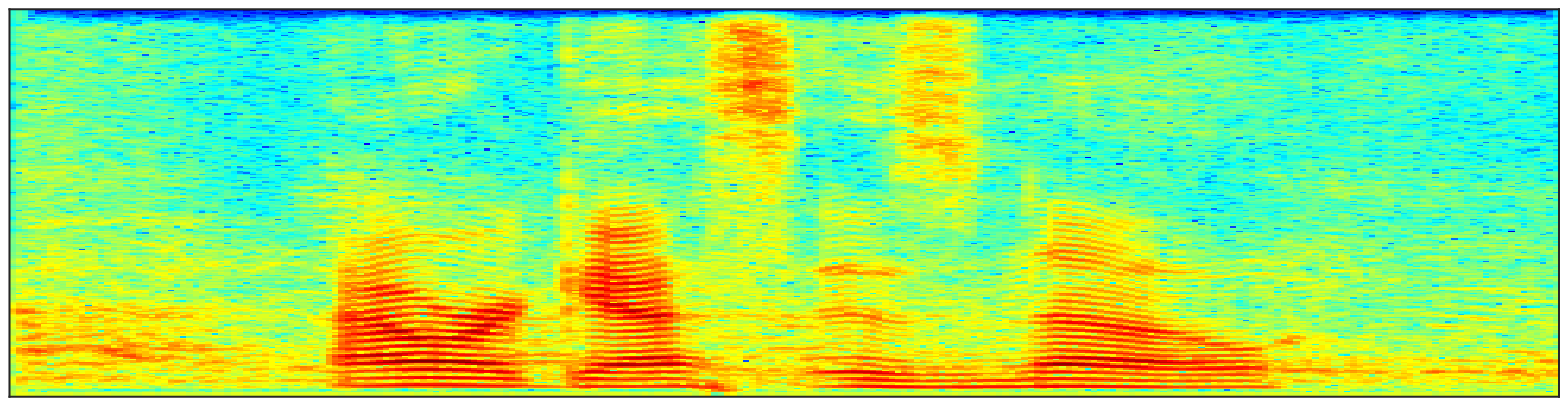};	
			\end{axis}
			\label{fig:illustration_diffuseness_comparison_S}
			\end{tikzpicture}
		}\\[-3mm]  
		\subfigure{
			\begin{tikzpicture}[scale=1,trim axis left]  
			\node at (-0.95,2.25) {\small (b)};
			\begin{axis}[
			label style = {font=\scriptsize},
			tick label style = {font=\scriptsize},   
			xlabel style={yshift=1mm},
			ylabel style={yshift=-1mm},
			width=\textwidth,height=3.85cm,grid=major,grid style = {dotted,black},
			axis on top, 	
			enlargelimits=false,
			xmin=1, xmax=245, ymin=0, ymax=8000,	
			xtick={1,50,100,150,200},
			xticklabels={\empty},
			change y base=true,  y SI prefix=kilo,
			ytick={0, 2000, 4000, 6000, 8000},
			ylabel={$f/\text{kHz} \rightarrow$},
			colorbar horizontal, colormap/jet, 
			colorbar style={
				at={(0,1.175)}, anchor=north west, font=\scriptsize, width=0.87\textwidth, height=0.25cm, xticklabel pos=upper
			},
			point meta min=0, point meta max=1]
			\addplot graphics [xmin=0, xmax=246, ymin=0, ymax=8025] {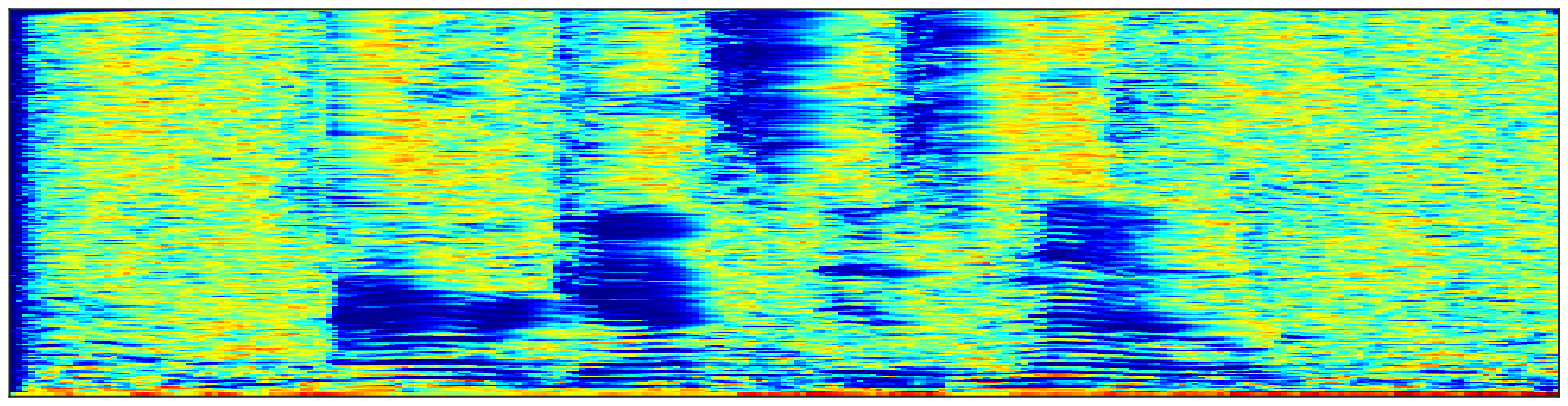};
			\end{axis}
			\label{fig:illustration_diffuseness_comparison_CDRprop3}
			\end{tikzpicture}
		}\\[-5mm]  
		\subfigure{
			\hspace*{-1.75mm}
			\begin{tikzpicture}[scale=1,trim axis left]  
			\node at (-0.95,2.25) {\small (c)};
			\begin{axis}[
			label style = {font=\scriptsize},
			tick label style = {font=\scriptsize},   
			xlabel style={yshift=1mm},
			ylabel style={yshift=-1mm},
			width=\textwidth,height=3.85cm,grid=major,grid style = {dotted,black},
			axis on top, 	
			enlargelimits=false,
			xmin=1, xmax=245, ymin=0, ymax=8000,	
			xtick={1,50,100,150,200},
			xlabel={$l\rightarrow$},
			change y base=true,  y SI prefix=kilo,
			ytick={0, 2000, 4000, 6000, 8000},
			ylabel={$f/\text{kHz} \rightarrow$}]
			\addplot graphics [xmin=0, xmax=246, ymin=0, ymax=8025] {diffuseness_CDRprop3_D_mic.eps};
			\end{axis}
			\label{fig:illustration_diffuseness_comparison_CDRprop2}
			\end{tikzpicture}
		}
		\caption{Comparison of the behaviour of the two different \ac{CDR} estimators in (\ref{eq:CDR_estimator_prop2}) and (\ref{eq:CDR_estimator_prop3}). Figure (a) shows the spectrogram of the recorded close-talking signal $20\log_{10}(|S(l,f)|)$. Figures (b) and (c) show the average diffuseness, estimated using the \ac{DOA}-independent and \ac{DOA}-dependent estimator in (\ref{eq:CDR_estimator_prop2}) and (\ref{eq:CDR_estimator_prop3}), respectively.}
		\label{fig:illustration_diffuseness_comparison}
	\end{figure}

	%
	In Figure~\ref{fig:illustration_diffuseness_comparison}, we illustrate the estimated diffuseness $\overline{\mathrm{D}}(l,f)$ obtained with the \ac{DOA}-dependent estimator (\ref{eq:CDR_estimator_prop3}) in Figure~\ref{fig:illustration_diffuseness_comparison_CDRprop3} and with the \ac{DOA}-independent estimator (\ref{eq:CDR_estimator_prop2}) in Figure~\ref{fig:illustration_diffuseness_comparison_CDRprop2}, to highlight the difference between our two estimators. As a reference, the spectrogram of the reference signal is shown in Figure~\ref{fig:illustration_diffuseness_comparison_S}, again. It can be seen that both estimators yield an estimated diffuseness with a similar structure, i.e., the diffuseness is of low value when the target signal is active and vice versa. However, it can be seen that the \ac{DOA}-dependent estimator in general attributes higher diffuseness values to time-frequency regions where the target signal is not active. 
	%
	%
	The reason for this difference lies in the fact that the \ac{DOA}-dependent estimator exhibits a directional filtering effect, whereas the \ac{DOA}-independent estimator does not. The DOA-dependent estimator only considers signal components arriving from the given target \ac{DOA} as desired signal components, while directional signal components from other directions will increase the diffuseness estimate. Consequently, the \ac{DOA}-dependent estimator will lead to stronger suppression in time-frequency regions where the target signal is not active.
	Even if no directional interferers are present, the \ac{DOA}-dependent estimator can achieve higher suppression of diffuse noise due to lower sensitivity to the variance of the coherence estimate for diffuse signal components \cite{zheng_IWAENC2016}.
	
	
	\subsection{Evaluation of recognition accuracy}
\begin{figure}
		\centering
		\begin{tikzpicture}[scale=1]
		\scriptsize
		\def\lx{0}
		\def\ly{3.2}
		\begin{axis}[
		width=\textwidth,height=4.5cm,grid=major,grid style = {dotted,black},
		ylabel={$\text{WER} \;/$ \;\% $\; \rightarrow$},
		xlabel={Overestimation factor $\mu$ $\; \rightarrow$},
		ymin=5.3,ymax=6.6,xmin=0.3,xmax=1.4,
		legend style={at={(0.5,1.375)},anchor=north,legend columns=3,font=\scriptsize, /tikz/every even column/.append style={column sep=10pt}},
		]
		\addplot[thick,black,solid] table [x index=0, y index=1]{NoPF.dat}; \addlegendentry{No WF}
		\addplot[thick,blue,solid,mark=*] table [x index=0, y index=1]{PFschwarz_noDOA_eval.dat}; \addlegendentry{WF\textsubscript{DOAindep}}
		\addplot[thick,green,solid,mark=square] table [x index=0, y index=1]{PFschwarz_DOA_eval.dat}; \addlegendentry{WF\textsubscript{DOAdep}}
		\addplot[thick,orange,solid,mark=+] table [x index=0, y index=1]{PFjeub_eval.dat}; \addlegendentry{WF\textsubscript{Jeub}}
		\addplot[thick,red,solid,mark=o] table [x index=0, y index=1]{PFthiergart_eval.dat}; \addlegendentry{WF\textsubscript{Thiergart}}
		\end{axis}
		\end{tikzpicture}
		\caption{\acp{WER} obtained by the \ac{CHiME-3} \ac{ASR} system for real recordings of the development set without (``No WF'') and with  coherence-based postfiltering of the beamformer output signal using \ac{DOA}-independent~(WF\textsubscript{DOAindep} and WF\textsubscript{Thiergart}) and \ac{DOA}-dependent (WF\textsubscript{DOAdep} and WF\textsubscript{Jeub}) \ac{CDR} estimation.}
		\label{fig:ResL}
	\end{figure}
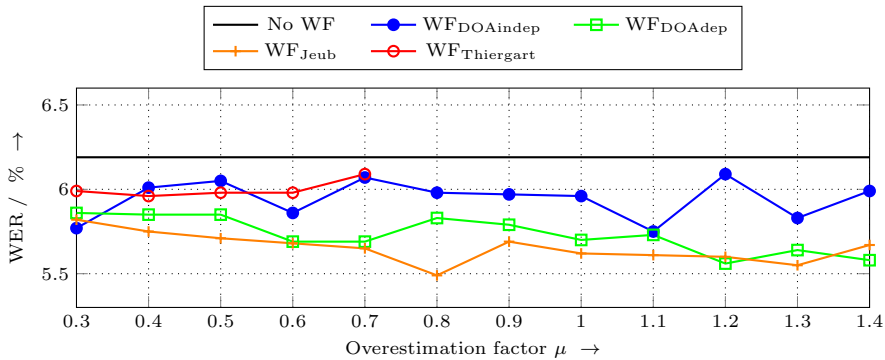

	In the following, we evaluate the recognition accuracy obtained by the front-end signal enhancement in Section~\ref{sec:FrontEnd} combined with the \ac{ASR} system in Section~\ref{sec:BackEnd}. More precisely, we evaluate the impact of coherence-based single-channel Wiener filtering, realized using ...
	\begin{itemize}
		\item \ac{DOA}-independent \ac{CDR} estimation (\ref{eq:CDR_estimator_prop2}), termed WF\textsubscript{DOAindep},
		\item \ac{DOA}-dependent \ac{CDR} estimation (\ref{eq:CDR_estimator_prop3}), termed WF\textsubscript{DOAdep},
		\item \ac{DOA}-independent \ac{CDR} estimation (\ref{eq:CDR_estimator_thiergart2}), termed WF\textsubscript{Thiergart},
		\item \ac{DOA}-dependent \ac{CDR} estimation (\ref{eq:CDR_estimator_jeub}), termed WF\textsubscript{Jeub}.
	\end{itemize}
	In Figure~\ref{fig:ResL}, the resulting \acp{WER} for real recordings of the \ac{CHiME-3} development set are illustrated for different values of the overestimation factor $\mu$ in (\ref{eq:WienerFilter}). 
	It is obvious that the recognition accuracy of the \ac{CHiME-3} ASR system without postfilter (``No WF'' in Figure~\ref{fig:ResL}) is consistently improved by incorporating coherence-based postfiltering %
	, except for WF\textsubscript{Thiergart} with $\mu \geq 0.8$ which yielded \acp{WER} between $6.45\%$ and $7.94\%$. Since these \acp{WER} are higher than those without postfiltering, they are not shown in Figure~\ref{fig:ResL} to allow for a clearer presentation of the remaining results.
	Moreover, it can be seen from Figure~\ref{fig:ResL} that the Wiener filters which were realized using \ac{DOA}-dependent \ac{CDR} estimators (WF\textsubscript{DOAdep} and WF\textsubscript{Jeub}) yield a better signal enhancement performance compared to Wiener filters realized using \ac{DOA}-independent \ac{CDR} estimators (WF\textsubscript{DOAindep} and WF\textsubscript{Thiergart}).
	Among the \ac{DOA}-dependent Wiener filter realizations, WF\textsubscript{Jeub} yields a slightly better performance than WF\textsubscript{DOAdep} for almost every value of $\mu$. 
	
	Based on the results illustrated in Figure~\ref{fig:ResL}, we selected one overestimation factor $\mu_\text{opt}$ for each Wiener filter realization, in order to obtain the best performance in the experiments using the evaluation data set (real data). The selected overestimation factors for each Wiener filter realization are summarized in Table~\ref{tab:table_best_mu}. 
	One can see that the overestimation factors chosen for WF\textsubscript{Thiergart} and WF\textsubscript{Jeub} are much smaller than those for WF\textsubscript{DOAindep} and WF\textsubscript{DOAdep}. To verify that there is no unfair comparison of the different Wiener filter realizations due to the potentially lower noise suppression performance, we also evaluated the real recordings of the evaluation data set for the different overestimation factors. Our results showed that larger overestimation factors as the ones in Table~\ref{tab:table_best_mu} for WF\textsubscript{Thiergart} and WF\textsubscript{Jeub} do not lead to better \acp{WER} for these Wiener filter realizations.   
	\begin{table}[t]
		\begin{center}    
			\caption{Selected overestimation factors $\mu_\text{opt}$ for which each Wiener filter realization achieved the best signal enhancement performance for the real recordings of the \ac{CHiME-3} development data set.}
			\label{tab:table_best_mu}   
			\begin{tabular}{c c c c c}
				\toprule   
				& WF\textsubscript{DOAindep} & WF\textsubscript{DOAdep} & WF\textsubscript{Thiergart} & WF\textsubscript{Jeub}\\
				\midrule
				$\mu_\text{opt}$ & $1.1$ & $1.2$ & $0.4$ & $0.8$\\
				\bottomrule    
			\end{tabular}
		\end{center}
	\end{table}
	
	In Table~\ref{tab:Results_avgWERs}, the average \acp{WER} (averaged over all environment-specific \acp{WER}) for the \ac{CHiME-3} development and evaluation set (both real data) obtained with the \ac{CHiME-3} \ac{ASR} system without (``No WF'') and with coherence-based postfiltering of the beamformer output signal using \ac{DOA}-independent (WF\textsubscript{DOAindep} and WF\textsubscript{Thiergart}) or \ac{DOA}-dependent (WF\textsubscript{DOAdep} and WF\textsubscript{Jeub}) \ac{CDR} estimation are compared. Moreover, the relative \ac{WER} improvements with respect to the baseline signal enhancement (without postfilter) are given.
	The results show that the coherence-based postfiltering of the beamformer output signal consistently improves the recognition accuracy of the \ac{ASR} system for both, the development set as well as the evaluation set. 
	For the development set, WF\textsubscript{Jeub} yields the lowest \ac{WER}, closely followed by WF\textsubscript{DOAdep}, with relative \ac{WER} improvements with respect to the baseline signal enhancement of $11.31\%$ and $10.18\%$, respectively. On the contrary, for the evaluation data set WF\textsubscript{DOAdep} performs best, followed by WF\textsubscript{DOAindep}. Here, the average \ac{WER} of the baseline signal enhancement is reduced by up to $8.21\%$.	
	
	To provide more insight into the behavior of the various Wiener filter realizations in the different acoustic environments provided by \ac{CHiME-3}, we provide the environment-specific \acp{WER} obtained for the evaluation set (real data) in Table~\ref{tab:Results_MVDR+PF_scenarioSpecificWERs}.
	We observe that except for the STR environment, WF\textsubscript{DOAdep} always yields the best results. In the STR environment, WF\textsubscript{DOAindep} proves to be slightly better. The latter might be due to incorrectly localized target source directions in this environment.
	%
	\begin{table*}
		\begin{center}    
			\caption{Average \acp{WER} and relative \ac{WER} improvements (in \%) for the \ac{CHiME-3} development and evaluation set (both real data) obtained with the \ac{CHiME-3} \ac{ASR} system without (``No WF'') and with coherence-based postfiltering of the beamformer output signal using \ac{DOA}-independent (WF\textsubscript{DOAindep} and WF\textsubscript{Thiergart}) and \ac{DOA}-dependent (WF\textsubscript{DOAdep} and WF\textsubscript{Jeub}) \ac{CDR} estimation. The relative \ac{WER} improvements are given with respect to the \ac{WER} of the baseline signal enhancement.}\vspace{2mm}
			\label{tab:Results_avgWERs}
			\begin{tabular}{c c c c c}
				\toprule   
				& \multicolumn{2}{c}{Development set} & \multicolumn{2}{c}{Evaluation set}\\
				& avg. \ac{WER} & rel. Impr. & avg. \ac{WER} & rel. Impr. \\
				\midrule
				No WF & 6.19 & - & 12.67 & - \\
				WF\textsubscript{DOAindep} & 5.75 & 7.10 & 11.78 & 7.02\\
				WF\textsubscript{DOAdep} & 5.56 & 10.18 & \bf{11.63} & \bfseries{8.21}\\
				WF\textsubscript{Thiergart} & 5.96 & 3.71 & 12.16 & 4.03\\ 
				WF\textsubscript{Jeub} & \bf{5.49} & \bfseries{11.31} & 12.13 & 4.26 \\
				\bottomrule    
			\end{tabular}
		\end{center}
		\begin{center}
			\caption{Environment-specific \acp{WER} (in \%) for the \ac{CHiME-3} evaluation set (real data) obtained with the \ac{CHiME-3} \ac{ASR} system without (``No WF'') and with coherence-based postfiltering of the beamformer output signal using \ac{DOA}-independent (WF\textsubscript{DOAindep} and WF\textsubscript{Thiergart}) and \ac{DOA}-dependent (WF\textsubscript{DOAdep} and WF\textsubscript{Jeub}) \ac{CDR} estimation.}\vspace{2mm}
			\label{tab:Results_MVDR+PF_scenarioSpecificWERs}
			\begin{tabular}{c c c c c}
				\toprule   
				& \multicolumn{4}{c}{Evaluation set}\\				
				& BUS      & CAF      & PED     & STR \\
				\midrule
				No WF & 18.53 & 11.39 & 10.50 & 10.27\\
				WF\textsubscript{DOAindep} & 17.71 & 9.71 & 9.79 & \bf{9.88}\\
				WF\textsubscript{DOAdep} & \bf{17.42} & \bf{9.36} & \bf{9.57} & 10.18\\
				WF\textsubscript{Thiergart} & 18.70 & 10.07 & 9.98 & 9.90\\
				WF\textsubscript{Jeub} & 18.47 & 9.73 & 10.05 & 10.25\\
				\bottomrule    
			\end{tabular}
		\end{center}
	\end{table*}

	\begin{table}
		\begin{center}
			\caption{Comparison matrix showing results of the \acs{MAPSSWE} test applied to the results of the various front-end signal enhancement algorithms for the evaluation set (real data). For the \acs{MAPSSWE} test a significance level of $p=5\%$ was used.}\vspace{2mm}
			\label{tab:Results_MAPSSWE}
			\begin{tabular}{l c c c c}
				\toprule   
				&  WF\textsubscript{DOAindep} &  WF\textsubscript{DOAdep} &  WF\textsubscript{Thiergart} &  WF\textsubscript{Jeub} \\				
				\midrule
				No WF &  WF\textsubscript{DOAindep} &  WF\textsubscript{DOAdep}
				&  WF\textsubscript{Thiergart} & WF\textsubscript{Jeub} \\
				\midrule
				WF\textsubscript{DOAindep} & & same & same & same \\
				\midrule
				WF\textsubscript{DOAdep} & same & & WF\textsubscript{DOAdep} & WF\textsubscript{DOAdep} \\	
				\midrule
				WF\textsubscript{Thiergart} & same & WF\textsubscript{DOAdep} & & same \\
				\midrule
				WF\textsubscript{Jeub} & same & WF\textsubscript{DOAdep} & same & \\				
				\bottomrule    
			\end{tabular}
		\end{center}
	\end{table}
	In order to investigate the statistical significance of the \ac{WER} improvements obtained with the various Wiener filter realizations for the real evaluation data set, we applied the \ac{MAPSSWE} test to the results in Table~\ref{tab:Results_avgWERs}. The \ac{MAPSSWE} test uses knowledge of aligned reference and hypothesis (produced by the \ac{ASR} system) sentence strings to locate segments within the sentence strings which contain misclassified content. To compare two different systems, the number of errors in each segment for each system is computed, and the null hypothesis that the mean difference in the number of word errors per segment between the two systems is zero is tested, see, e.g. \cite{gillick:icassp1989,pallet:icassp1990}. In our experiments, a significance level of $p=5\%$, i.e., a $95\%$ confidence level for rejecting the null hypothesis, was chosen. We used the implementation of the \ac{MAPSSWE} test provided by the \ac{NIST} Scoring Toolkit \cite{sctk:2016}. Table~\ref{tab:Results_MAPSSWE} compares the results of the \ac{MAPSSWE} test. For every comparison, the better signal enhancement system is given in the corresponding field of the table if a statistically significant difference between the systems was found, and ``same'' otherwise.		
	All Wiener filter realizations yield a statistically significant improvement of the signal enhancement baseline without postfiltering of the beamformer output signal. It can furthermore be observed that WF\textsubscript{DOAdep} performs significantly better than WF\textsubscript{Thiergart} and WF\textsubscript{Jeub}, but no significant difference between WF\textsubscript{DOAdep} and WF\textsubscript{DOAindep} is found.
	Thus, a consistent and statistically significant benefit of exploiting \ac{DOA} information for \ac{CDR}-based postfiltering cannot be inferred from the results. An additional significance test for the results of the different Wiener filter realizations for the real development data set did also not show a consistent and statistically significant advantage of \ac{DOA}-dependent over \ac{DOA}-independent \ac{CDR}-estimators.
	%
	%
	%
	\section{Conclusion}
	\label{sec:summary_conclusion}
	We proposed to extend the front-end speech enhancement of a state-of-the-art \ac{ASR} system by coherence-based postfiltering of the beamformer output signal. The postfilter is realized as a Wiener filter, where an estimate of the power ratio between direct and diffuse signal components at the output of the beamformer is used as an approximation of the short-time \ac{SNR} to compute the time- and frequency-dependent postfilter gains. To estimate the ratio between direct and diffuse signal components, we used two \ac{DOA}-independent and two \ac{DOA}-dependent estimators, which can be efficiently realized by estimating the auto- and cross-\ac{PSDs} at the microphone signals. As a consequence, the postfilter has a very low computational complexity.
	Baseline and extended front-end speech enhancement have been evaluated on real recordings provided by \ac{CHiME-3} with respect to \acp{WER} of a state-of-the-art \ac{HMM}-\ac{DNN}-based \ac{ASR} system. The results confirmed that coherence-based postfiltering in general improves the recognition accuracy of the \ac{ASR} system significantly, with relative improvements of up to $11.31\%$ and $8.21\%$ for the development and the evaluation data set, respectively.
	Consistent statistically significant differences between Wiener filters based on \ac{DOA}-independent and \ac{DOA}-dependent \ac{CDR} estimators could not be observed. 
	The improved recognition accuracy in addition to the low computational complexity makes coherence-based postfiltering very suitable for real-time robust distant speech recognition.  
	As future work, it should be evaluated whether \ac{CDR}-based postfilters for dereverberation still yield a significant improvement of \acp{WER} when a more powerful beamforming algorithm is used: On the one hand, if the beamformer has a higher directivity, i.e., a higher suppression of diffuse noise components, a CDR-based postfilter might be less effective because the beamformer partly fulfills the function of the \ac{CDR}-based postfilter. On the other hand, if the beamformer already partly suppresses the diffuse noise, then the postfilter sees a better input \ac{CDR} and could possibly be tuned to be more aggressive, which could further improve speech recognition performance. Another aspect of future work is the investigation of the combination of \ac{DOA}-independent and \ac{DOA}-dependent \ac{CDR} estimators in different frequency areas, in order to obtain an even better diffuseness estimate.

	\section{Acknowledgement}
	\label{sec:acknowledgements}
	We would like to thank Stefan Meier and Christian Hofmann for their continuous support and fruitful discussions.
	
	The research leading to these results has received funding from the European Union's Seventh Framework Programme (FP7/2007-2013) under grant agreement n$^\mathsf{o}$ 609465 and from the Deutsche Forschungsgemeinschaft (DFG) under contract number KE 890/4-2.
	
	\section*{References}
	\bibliography{elsevier_multimicasr2016}
	
\end{document}